\documentstyle[12pt,epsfig]{article}

\begin{document}
\baselineskip=23pt

\vspace{1.2cm}

\begin{center}
{\huge \bf  Unification of Gravitation and Gauge Fields}

\bigskip

Xin-Bing Huang\footnote{huangxb@mail.phy.pku.edu.cn}\\
{\em Department of Physics,
Peking University,} \\
{\em  100871 Beijing, China}
\end{center}

\bigskip
\bigskip
\bigskip

\centerline{\large Abstract} In this letter, I indicate that
complex daor field should also have spinor suffixes. The
gravitation and gauge fields are unified under the framework of
daor field. I acquire the elegant coupling equation of gravitation
and gauge fields, from which Einstein's gravitational equation can
be deduced.

\vspace{1.2cm}

 PACS numbers: 04.20.Cv, 04.20.Gz, 03.65.Vf, 12.10.Kt

\vspace{1.2cm}

\newpage

About four decade years ago, some physicists recognized the fact
that Yang-Mills gauge theories and the affine geometry of
principal fiber bundles are one\cite{egf80,yan75}. But Einstein's
gravitational theory is the affine geometry of tangent bundles.
They seems to be quite different. I have indicated that daor field
will construct a possible connection between them\cite{hua04}.
This letter is devoted to this topic: Gravitation and gauge field
are unified in a harmonic structure, and the coupling equation is
set up, which is consistent with Einstein's gravitational theory.

Suppose an ideal universe, in which there is no matter present
except the gravitational field and gauge fields. Einstein's
 gravitational equation can be written as\footnote{In this letter, using Roman suffixes to
refer to the bases of local Minkowski frame; using Greek suffixes
to refer to curvilinear coordinates of
space-time.}\cite{mtw73,dir75}
\begin{equation}
\label{daor1} G_{\mu\nu}=R_{\mu\nu}-\frac{1}{2} ~ g_{\mu\nu}R=8\pi
G T_{\mu\nu} ~,
\end{equation}
where $G$ is Newtonian gravitational constant, and $T_{\mu\nu}$ is
the stress-energy tensor for gauge fields. In the case of
electromagnetic field, $T_{\mu\nu}$ is given by
\begin{equation}
\label{daor101} 4\pi T_{\mu\nu}={\bf f}_{\mu}^{~\alpha} {\bf
f}_{\nu\alpha} -\frac{1}{4} ~ g_{\mu\nu}{\bf f}_{\alpha\beta}{\bf
f}^{\alpha\beta}~,
\end{equation}
where ${\bf f}_{\alpha\beta}$ is the strength of electromagnetic
field. I adopt the same sign conventions as in
Misner-Thorne-Wheeler's book\cite{mtw73}. The metric tensor of
Minkowski space-time $\eta_{ab}$ is written as follows
\begin{equation}
\label{daor102}
\eta^{00}=-1~,~~~~\eta^{11}=\eta^{22}=\eta^{33}=+1~,~~~~
\eta^{ab}=0~~~~{\rm for}~~~~a \neq b~.
\end{equation}

In Minkowski space-time, Dirac equation is usually written
as($\hbar=c=1$)\cite{dir58}
\begin{equation}
\label{daor2} \left(i\gamma^{a}\frac{\partial}{\partial
x^{a}}-m\right)\psi=0~.~~~~~~a=0,1,2,3.
 \end{equation}
Where $\gamma$'s are Dirac matrices, which satisfy
\begin{equation}
\label{daor201}
\gamma^{a}\gamma^{b}+\gamma^{b}\gamma^{a}=-2\eta^{ab}~.
\end{equation}

In my former paper\cite{hua04}, I had given the concept of daor
field, which can be regarded as the square root of space-time
metric. Daor field $h^{a}_{~\mu}$ or $H^{~\mu}_{a}$ satisfies
\begin{equation}
\label{daor4} g_{\mu\nu}=h^{*~a}_{\mu}\eta_{ab}h^{b}_{~\nu}~
,~~~~~~G^{\mu\nu}=H^{*\mu}_{~~a}\eta^{ab}H^{~\nu}_{b}~,~~~~~~
g_{\mu\nu}G^{\nu\lambda}=G^{\lambda\nu}g_{\nu\mu}=\delta^{\lambda}_{\mu}~.
\end{equation}
where $*$ denotes complex conjugation. Set $h^{a}=h^{a}_{~\mu}{\rm
d}x^{\mu}$, which is daor field 1-form. By defining the Hermitean
conjugate of daor field $h^{a}_{~\mu}$ or $H^{~\mu}_{a}$ as
follows
\begin{equation}
\label{daor5}
(h^{a}_{~\mu})^{\dag}=h^{*~a}_{\mu}~,~~~~~~(H^{~\mu}_{a})^{\dag}=H^{*\mu}_{~~a}~,
\end{equation}
we can easily acquire the following relations
\begin{equation}
\label{daor6} g=h^{\dag}\eta h~ ,~~~~G=H^{\dag}\eta H~,~~~~
G=g^{-1}~,~~~~H^{\dag}=h^{-1}~,~~~~H=(h^{\dag})^{-1}~.
\end{equation}

By using the same definition of inner product as in differential
geometry, the inner product of the vector
$U=U^{\mu}\frac{\partial}{\partial x^{\mu}}$ and the covector
$v=v_{\nu}{\rm d}x^{\nu}$ can be expressed as follows
\begin{equation}
\label{daor604}
<U,v>=<v,U>=U^{\mu}v_{\mu}=v_{\mu}U^{\mu}=U^{*a}v_{a}=v^{*}_{a}U^{a}~.
\end{equation}
Where $U^{*a}$, $U^{a}$, $v^{*}_{a}$ and $v_{a}$ are given by
\begin{equation}
\label{daor605}
U^{*a}=U^{\mu}h_{\mu}^{~*a}~,~~~~U^{a}=h_{~\mu}^{a}U^{\mu}~,
~~~~v^{*}_{a}=v_{\mu}H_{~~a}^{*\mu}~,~~~~v_{a}=H_{a}^{~\mu}v_{\mu}~.
\end{equation}
For simplicity, I will do not distinguish between the component
$U^{a}$, $v_{a}$ and its complex conjugate because they all can be
transferred into real component in curvilinear coordinates
$x^{\mu}$'s by corresponding form of daor field. In this daor
geometry, the exterior derivative and exterior product have the
same definitions and properties as in ordinary real vierbein
geometry.

Now let us discuss a kind of gauge groups, which are the subgroups
of U(1,3) group, and the element of gauge group can be written as
\begin{equation}
\label{daor7} S_{~b}^{a}(x)=e^{iZ^{a}_{~b}(x)}~.
\end{equation}
Here, $Z^{a}_{~b}$ is $4\times 4$ matrix, which is traceless and
Hermitean, of course, should be the function of curvilinear
coordinates. So $S^{a}_{~b}$ is the 4-dimensional representation
of gauge group. Some groups such as U(1), SU(2) and SU(3) all
satisfy the condition (\ref{daor7}).

An intrinsic rotation of daor field is
\begin{equation}
\label{daor15} h^{a}\rightarrow h^{\prime a}=S^{a}_{~b}h^{b}~,
\end{equation}
here $S^{a}_{~b}$ satisfies
\begin{equation}
\label{daor16} S^{*c}_{a}\eta_{cd}S^{d}_{~b}=\eta_{ab}~.
\end{equation}
From references\cite{hua04,egf80}, it is known that under the
intrinsic rotation of daor field the complex affine connection
1-form $\omega^{a}_{b}$ transforms as follows
\begin{equation}
\label{daor17} \omega^{\prime
a}_{~~b}=S^{a}_{~c}\omega^{c}_{~d}(S^{-1})^{d}_{~b}+S^{a}_{~c}({\rm
d }S^{-1})^{c}_{~b}~,
\end{equation}
Covariance of the daor field equations under local gauge group
directly leads to the introduction of Yang-Mills gauge
fields\cite{yan54}. Similarly, I separate gauge field
$B^{a}_{~b\mu}$ from the complex affine connection
$\omega^{a}_{~b\mu}$, say, write $\omega^{a}_{~b}$ as
\begin{equation}
\label{daor18} \omega^{ a}_{~b}=\Omega^{a}_{~b}+i
\epsilon^{\prime} B^{a}_{~b}~,
\end{equation}
where $\epsilon^{\prime}$ is the coupling constant of gauge field,
$B^{a}_{~b}$ is also 1-form. Under the gauge rotation of daor
field (\ref{daor15}), $B^{a}_{~b}$ transforms as follows
\begin{equation}
\label{daor19} B^{\prime
a}_{~~b}=S^{a}_{~c}B^{c}_{~d}(S^{-1})^{d}_{~b}+\frac{1}{\epsilon^{\prime}}
~ S^{a}_{~c}({\rm d
}S^{-1})^{c}_{~b}=S^{a}_{~c}B^{c}_{~d}(S^{-1})^{d}_{~b}+\frac{1}{\epsilon^{\prime}}
~{\rm d }Z^{c}_{~b}~,
\end{equation}
and $\Omega^{a}_{~b}$  satisfies
\begin{equation}
\label{daor1901} \Omega^{\prime
a}_{~~b}=S^{a}_{~c}\Omega^{c}_{~d}(S^{-1})^{d}_{~b}~.
\end{equation}
As having been given in the paper of Yang and Mills\cite{yan54},
the gauge field strengths corresponding to gauge field
$B^{a}_{~b}$ is given by\footnote{In this letter, {\rm d} and
$\wedge$ are exterior derivative and exterior product operators
respectively.}
\begin{equation}
\label{daor1902} F^{a}_{~b}={\rm d
}B^{a}_{~b}+i{\epsilon^{\prime}}B^{a}_{~c}\wedge B^{c}_{~b}~.
\end{equation}
Where $F^{a}_{~b}$ is a 2-form, which is a part of the curvature
2-form defined by
\begin{equation}
\label{daor1903} R^{a}_{~b}={\rm d
}\omega^{a}_{~b}+\omega^{a}_{~c}\wedge\omega^{c}_{~b}~.
\end{equation}
It is stressed that $F^{a}_{~b}$ has quite the same symmetric
characters as $R^{a}_{~b}$. It is well known that the
stress-energy tensor for this gauge field can be written as
\begin{equation}
\label{daor20} 4\pi T_{\mu\nu}={\rm tr}( F_{\mu}^{~\alpha}
F_{\nu\alpha}) -\frac{1}{2} ~ g_{\mu\nu}{\rm
tr}(F_{\alpha\beta}F^{\alpha\beta})~,
\end{equation}
Here `tr' means operation of acquiring the trace of a matrix. In
the following, I will give the coupling equation of daor field
with gauge fields when the stress-energy tensor in
Eq.(\ref{daor1}) is the form of Eq.(\ref{daor20}).

Yeah, I found that the doar-gauge field coupling equation can be
written as follows
\begin{equation}
\label{daor21} {\rm d
}h^{a}+(\omega^{a}_{~b}+i~\epsilon~\gamma^{c}~F^{a}_{~bc})\wedge~h^{b}=0~.
\end{equation}
Where $\epsilon$ is the coupling constant, which equal to the
square root of the newtonian gravitational constant, namely
$\epsilon=\sqrt{G}$. It should be noted that
$\gamma^{c}F^{a}_{~bc}$ is a 1-form also, what is to say
\begin{equation}
\label{daor22}
\gamma^{c}F^{a}_{~bc\mu}dx^{\mu}=\gamma^{\mu}F^{a}_{~b\mu\nu}dx^{\nu}~.
\end{equation}
Eq.(\ref{daor21}) also demonstrates that complex daor field should
also have spinor suffixes.

I will prove that Einstein's gravitational equation can be deduced
from Eq.(\ref{daor21}). Set $\epsilon=1$ in the process of proving
for simplicity. Firstly, define the operator 1-form
\begin{equation}
\label{daor23} \hat{W}\equiv\delta^{a}_{~b}{\rm d
}+(\omega^{a}_{~b}+i\gamma^{c}F^{a}_{~bc})\wedge~,
\end{equation}
then, Eq.(\ref{daor21}) becomes $\hat{W}h=0$. Multiplying both
sides of Eq.(\ref{daor21}) by operator $\hat{W}$, we acquire
\begin{eqnarray}
\label{daor24}
\nonumber
0&=&\hat{W} \hat{W} h
\\
\nonumber
 &=& \left[ {\rm d
}(\omega^{a}_{~b}+i\gamma^{c}F^{a}_{~bc})+(\omega^{a}_{~e}+i\gamma^{c}F^{a}_{~ec})\wedge
(\omega^{e}_{~b}+i\gamma^{c}F^{e}_{~bc}) \right] \wedge h^{b}
\\
 &=& \left[ R^{a}_{~b}+i\gamma^{c}({\rm d
}F^{a}_{~bc}+\omega^{a}_{~e}\wedge
F^{e}_{~bc}+F^{a}_{~ec}\wedge\omega^{e}_{~b})-\gamma^{c}\gamma^{d}F^{a}_{~ec}\wedge
F^{e}_{~db} \right] \wedge h^{b}~.
\end{eqnarray}
The covariant derivative of a differential form $V^{a}_{~b}$ of
degree $p$ is defined as\cite{egf80}
\begin{equation}
\label{daor25} DV^{a}_{~b}={\rm d } V^{a}_{~b}+\omega^{a}_{~c}
\wedge V^{c}_{~b}- (-1)^{p} V^{a}_{~c} \wedge \omega^{c}_{~b}~.
\end{equation}
Because $F^{a}_{b}$ is the gauge field strength, it satisfies the
following Bianchi identities:
\begin{equation}
\label{daor26} DF^{a}_{~b}=0~.
\end{equation}
As I have stressed that $\gamma^{c}F^{a}_{~bc}$ is a 1-form,
Eq.(\ref{daor25}) and Eq.(\ref{daor26}) then make sure that
$\gamma^{c}({\rm d }F^{a}_{~bc}+\omega^{a}_{~e}\wedge
F^{e}_{~bc}+F^{a}_{~ec}\wedge\omega^{e}_{~b})=0$. So
Eq.(\ref{daor24}) becomes
\begin{equation}
\label{daor2601} R^{a}_{~b}=\gamma^{c}\gamma^{d}F^{a}_{~ec}\wedge
F^{e}_{~db}~.
\end{equation}
Transferring $R^{a}_{~b}$ into the curvilinear coordinates of
space-time, from Eq.(\ref{daor2601}) we obtain
\begin{eqnarray}
\label{daor27}
R^{\alpha}_{~\beta\mu\nu}=-\gamma^{c}\gamma^{d}(F^{\alpha}_{~ec\nu}
F^{e}_{~\beta d \mu}-F^{e}_{~\beta d \mu}F^{\alpha}_{~ec\nu})
\end{eqnarray}
Let us contract the suffixes $\alpha$ and $\mu$ in
$R^{\alpha}_{~\beta\mu\nu}$
\begin{eqnarray}
\label{daor29}\nonumber
R_{\beta\nu}&=&-\sum_{\alpha}F^{\alpha}_{~ec\nu} F^{e}_{~\beta d
\alpha}(\gamma^{c}\gamma^{d}+\gamma^{d}\gamma^{c}) \\
\nonumber
&=&2\sum_{a}F^{a}_{~e\lambda\nu} F^{e~\lambda}_{~\beta~~  a} \\
&=&2{\rm tr}(F_{\lambda\nu}F^{\lambda}_{~\beta})~.
\end{eqnarray}
Resuming the value of $\epsilon$, we can acquire Einstein's
gravitational equation (\ref{daor1}) and the stress-energy tensor
(\ref{daor20}).

When there are different categories of gauge fields in the
space-time, the coupling equation can be extended as follows
\begin{equation}
\label{daor30} {\rm d
}h^{a}+(\omega^{a}_{~b}+i~\epsilon~\gamma^{c}~F^{a}_{~bc})\wedge~h^{b}=0
~,~~~~~~F^{a}_{~b\mu\nu}=\sum_{\tau}~^{\tau}F^{a}_{~b\mu\nu}~,
\end{equation}
where $^{\tau}F^{a}_{~b\mu\nu}$ denotes the strength of different
gauge fields.

 The coupling
equation (\ref{daor21}) demonstrates that the coupling constant
$\epsilon$ is unrelated with the category of gauge field. This
reflects the generality of gravitation. I believe, the coupling
constant between daor field and spinor field should also be
$\epsilon$. The coupling between daor field and spinor field will
be discussed in forthcoming papers.

The coupling equation (\ref{daor21}) also indicates that only daor
field can express the intrinsic harmony of different fields.

Conclusion: The general form of gauge fields are discussed. All
gauge fields originate the invariance of local intrinsic rotation
of doar field. The coupling equation is submitted, from which
Einstein's equation can be obtained.

 {\bf Acknowledgement:}  I would like to thank my parents , Long-Mei Zhu
 and You-Ping Dai for their help and encouragement.

\end{document}